\documentclass[pre,a4paper,onecolumn,noshowpacs,noshowkeys,nofootinbib,floatfix]{revtex4}
\usepackage{amsfonts}
\usepackage{amsmath}
\usepackage{amssymb}
\usepackage{graphicx}

\begin{document}

\title{On a generalised model for time-dependent variance with long-term memory}
\author{S.~M.~Duarte~Queir\'{o}s}
\email[e-mail address: ]{sdqueiro@cbpf.br}
\affiliation{Centro Brasileiro de Pesquisas F\'{\i}sicas, 150, 22290-180, Rio de Janeiro
- RJ, Brazil}
\date{\today}

\begin{abstract}
The ARCH process (\textsc{R. F. Engle}, $1982$) constitutes a paradigmatic
generator of stochastic time series with time-dependent variance like it
appears on a wide broad of systems besides economics in which ARCH was
born. Although the ARCH process captures the so-called \textquotedblleft
volatility clustering\textquotedblright\ and the asymptotic power-law
probability density distribution of the random variable, it is not capable
to reproduce further statistical properties of many of these time series
such as: the strong persistence of the instantaneous variance characterised
by large values of the Hurst exponent ($H > 0.8$), and asymptotic
power-law decay of the absolute values self-correlation function. By means of considering an
effective return obtained from a correlation of past returns that has a $q$%
-exponential form ($\exp _{q}\left[ x\right] \equiv \left[ 1+\left(
1-q\right) \,x\right] ^{\frac{1}{1-q}}$, $\left( q\in \Re \right) $, and 
$\exp _{1}\left[ x\right] =e^{x}$) we are able to fix the
limitations of the original model. Moreover, this improvement can be obtained through the correct choice of a sole additional parameter, $q_{m}$. 
The assessment of its validity and usefulness is made by mimicking daily fluctuations of $SP500$ financial index.
\end{abstract}

\maketitle

The time evolution analysis of both physical and non-physical observables
plays a central role in nowadays scientific research devoted to complexity.
Explicitly, studies on time series with geophysical, meteorological,
physiological, and financial origin, amid others, have populated scientific
literature particularly during the last two decades~\cite{complexity}.
Although each type of system has its own (microscopic) dynamical
mechanism, the fact is that certain time series obtained from so dispair
systems, like those mentioned above, exhibit common statistical features such
as \emph{asymptotic power-law decaying probability density functions}, and a 
\emph{long-lasting power-law-like self-correlation function of the magnitude
of the time series observable}, notwithstanding a fast vanishing or null
self-correlation function of the variable itself. This especial class of
time series has usually been associated with stochastic processes which have
time-dependent variance, mathematically defined as \textit{heteroskedasticity%
}, in contrast to the other type of time series, said 
\textit{homoskedastic}, that present a constant value for the variance.
Customarily, the profile of heteroskedastic time series is also reminiscent of 
\emph{on-off intermittency}~\cite{on-off}, {\it i.e.}, large values of the
variable upon analysis are typically followed by other large values, but
with an arbitrary sign. Within a financial context, this behaviour has been
found in price fluctuations of stocks traded in financial markets or
inflation~\cite{econofisica}. In $1982$, to further mimic the latter, 
\textsc{R.F. Engle} introduced the \emph{autoregressive conditional
heteroskedasticity} ($ARCH$) process~\cite{engle}. This process is
considered as a cornerstone of econometrics fact that awarded \textsc{Engle}
the $2003$ Nobel Memorial Prize in Economics ``for methods of analysing economic
time series with time-varying volatility''~\footnote{%
Volatility is the financial technical term for instantaneous variance.}.
Albeit its attested expressed in its broad application and generalisations~\cite{arch-rev}, 
\textsc{Engle}'s $ARCH$\ process is unable to properly reproduce the
long-lasting behaviour of the volatility self-correlation function, because
it only leads to an exponential decay of this function~\cite{boller}. In the
sequel of this manuscript we propose a generalisation of the celebrated $%
ARCH\left( 1\right) $ process by introducing a memory kernel emerging from
current non-extensive statistical mechanics formalism~\cite{GM-CT}. As a
result, we are able to obtain the asymptotic power-law behaviour of the
random variable probability density function, which is notably described by $%
q$-Gaussian distributions, and to mend the shortcoming of \textsc{Engle}'s
process. The usefulness of this generalisation is shown by modelling daily
fluctuations of $SP500$ financial index. 
%In addition, our model introduces a
%new triplet of entropic indices from non-extensive statistical mechanics 
%\cite{tsallis-villa}, which are related to the memory of the system, $q_{m}$%
%, the form of the probability density function, $q$, and the correlation
%function of the volatility, $q_{c}$.

\medskip

Following \textsc{Engle}~\cite{engle}, we define an autoregressive
conditional heteroskedastic ($ARCH$) time series as a discrete stochastic
process, $z_{t}$,
\begin{equation}
z_{t}=\sigma _{t}\ \omega _{t},  \label{arch-def}
\end{equation}
where $\omega _{t}$ is an independent and identically distributed random
variable with null mean and unitary variance, \textit{i.e.}, $\left\langle
\omega _{t}\right\rangle =0$ and $\left\langle \omega _{t}^{2}\right\rangle
=1$. Henceforth we call $z_{t}$ as \textit{return}. Normally, $\omega $ is
associated with a Gaussian distribution (which we have used throughout this
work), but other distributions for $\omega $ have been presented~\cite{noise-gen}. 
In the seminal paper of reference~\cite{engle}, it has been
suggested a possible dynamics for $\sigma _{t}^{2}$ (hereinafter denominated
as \textit{squared volatility}) defining it as a linear function of past
squared values of $z_{t}$,
\begin{equation}
\sigma _{t}^{2}=a+\sum \limits_{i=1}^{s}b_{i}\ z_{t-i}^{2},\qquad \left(
a,b_{i}\geq 0\right) .  \label{arch-vol}
\end{equation}
For its linear dependence on $z_{t-i}^{2}$, eq.~(\ref{arch-def}), together
with eq.~(\ref{arch-vol}), have been coined as $ARCH\left( s\right) $ \emph{%
linear process}. In financial practice, namely price fluctuation modelling,
the case $s=1$ ($b_{1}\equiv b$) is, by far, the most studied and applied of
all $ARCH$-like processes. It can be easily verified, even for all $s$,
that, although $\left\langle z_{t}\ z_{t^{\prime }}\right\rangle \sim \delta
_{t\,t^{\prime }}$, correlation $\left\langle \left\vert z_{t}\right\vert \
\left\vert z_{t^{\prime }}\right\vert \right\rangle $ is not proportional to 
$\delta _{t\,t^{\prime }}$. As a matter of fact, it has been proved for $s=1$
that, $\left\langle \left\vert z_{t}\right\vert \ \left\vert z_{t^{\prime
}}\right\vert \right\rangle $ decays as an exponential law with
characteristic time $\tau \equiv \left\vert \ln b\right\vert ^{-1}$, which
does not reproduce empirical evidences. In addition, it can be verified that, the process is stationary with a 
\textit{stationary variance}, $\bar{\sigma}$, 
\begin{equation}
\bar{\sigma}=\frac{a}{1-b}, \qquad (b>1),
\label{statvar}
\end{equation}
It has also been proved that, even
for large $s$, the exponential decay of $\left\langle \left\vert
z_{t}\right\vert \ \left\vert z_{t^{\prime }}\right\vert \right\rangle $
remains (check ref.~\cite{boller} for details). Furthermore, the
introduction of a large value for parameter $s$ gives rise to implementation
problems. In other words, when $s$ is large, it is very hard to find a set
of $\left\{ b_{i}\right\} $, since it represents the evaluation of a large
number of fitting parameters~\footnote{%
A generalisation of eq.~(\ref{arch-vol}), 
$\sigma _{t}^{2}=a+\sum \limits_{i=1}^{s}b_{i}\
z_{t-i}^{2}+\sum \limits_{i=1}^{r}c_{i}\ \sigma _{t-i}^{2}$ $\left(
a,b_{i},c_{i}\geq 0\right) $, known as $GARCH\left( s,r\right) $ process~\cite{granger}, 
was introduced in order to have a more flexible structure which
could correctly mimic data with a simple $GARCH\left( 1,1\right) $ process.
However, even this process presents an exponential decay for $\left\langle
\left\vert z_{t}\right\vert \ \left\vert z_{t^{\prime }}\right\vert
\right\rangle $, with $\tau \equiv \left\vert \ln \left( b+c\right)
\right\vert ^{-1}$, though condition $b+c<1$ guarantees that $GARCH\left(
1,1\right) $ corresponds exactly to an infinite-order $ARCH$ process.}. 
Despite instantaneous volatility fluctuation, the $ARCH(1)$ process is actually stationary and it presents 
a stationary returns probability density function with larger kurtosis than the distribution $P(\omega )$.
The kurtosis excess is precisely the outcome of such 
time-dependence of $\sigma _{t}$. Correspondingly, when $b=0$, the process reduces to generating a signal with 
the same PDF of $\omega $, but with a standard variation $\sqrt{a}$.

\medskip

We shall now introduce our variation on the $ARCH$ process. Explicitly, we
consider a $ARCH\left( 1\right) $ process where an effective immediate past
return, $\tilde{z}_{t-1}$, is assumed in the evaluation of $\sigma _{t}^{2}$%
. By this we mean that we have changed eq.~(\ref{arch-vol}) by 
\begin{equation}
\sigma _{t}^{2}=a+b\,\tilde{z}_{t-1}^{2},\qquad \left( a,b_{i}\geq 0\right) ,
\label{vol-qarch}
\end{equation}%
in which the effective past return is calculated according to%
\begin{equation}
\tilde{z}_{t}^{2}=\sum \limits_{i=t_{0}}^{t}\mathcal{K}\left( i-t\right)
\,z_{i}^{2},\qquad \left( t_{0}\leq t\right) ,
\end{equation}
where
\begin{equation}
\mathcal{K}\left( t^{\prime }\right) =\frac{1}{\mathcal{Z}_{q_{m}}\left(
t^{\prime }\right) }\exp _{q_{m}}\left[ t^{\prime } \right] ,\qquad
\left( t^{\prime }\leq 0,q_{m}<2\right)   \label{kernel}
\end{equation}%
with 
\begin{equation}
\exp _{q}\left[ x\right] \equiv \left[ 1+\left( 1-q\right) \,x\right] _{+}^{%
\frac{1}{1-q}},  \label{q-exp}
\end{equation}%
$\mathcal{Z}_{q_{m}}\left( t^{\prime }\right) \equiv \sum_{i=-t^{\prime
}}^{0}\exp _{q_{m}}\left[ i \right] $ ($\left[ x\right] _{+}=\max
\left\{ 0,x\right\} $~\footnote{%
This condition is known in the literature as \textit{Tsallis cut off} at $%
x = \left( 1-q\right) ^{-1}$.}). For $q=-\infty $, we obtain the standard 
$ARCH\left( 1\right) $, and for $q=1$, we have $\mathcal{K}\left( t^{\prime
}\right) $ with an exponential form since $\exp _{1}\left[ x\right] =e^{x}$~\cite{GM-CT}. 
Although it has a non-normalisable kernel, let us refer that the value $q_{m}= \infty $ 
corresponds to the situation in which all past returns have the same weight, 
\mbox{$ \mathcal{K}\left( t^{\prime }\right)=1/(t-t_{0}+1).$} 
The introduction of an exponential kernel has already been made 
in~\cite{dose} but, as stated therein, it is not able to capture the
long-lasting correlation in $\sigma _{t}$ (or $\left\vert z_{t}\right\vert $%
), at least for financial markets~\footnote{%
A worth mentioning continuous time aproach to price dynamics in stock markets
using an exponential kernel was presented in ref.~\cite{borland}.}. Even though,
we surmise that some systems (apart those we aim to replicate herein) might
have a set of its statistical properties well-described by processes for which 
$q_{m}\leq 1$. Considering the process as stationary, it is not difficult to verify that eq.~(\ref{statvar}) holds.

Moving ahead on the study of our proposal we 
have performed numerical realisations, based on eq.~(\ref{arch-def}) and eq.~(%
\ref{vol-qarch}), from which we\ have analysed the return probability
density function (PDF), the Hurst exponent~\cite{feder} of $\left\vert z_{t}\right\vert $ 
integrated signal as well as the $\left\vert z_{t}\right\vert $ self-correlation
function. In order that our goal is to verify the usefulness of eq.~(\ref{kernel}) 
we have kept $a=\frac{1}{2}$. 
% and $T=1$ for all realisations and we
%have only modified \textquotedblleft leading\textquotedblright\ parameters $%
%q_{m}$ and $b$. 
Our option is justified by the fact that $a$ might be eliminated if we define a new variable, 
$z ^{\prime } \equiv z / \sqrt{a}$, for which standard deviation becomes equal to 1 (when $b = 0$). Besides, expanding eq.~(\ref{vol-qarch}), 
$$\sigma _{t} = \sqrt{1+b\,\tilde{z}_{t}^{\prime \, 2}} \, \sim \, 1+\frac{b\,\tilde{z}_{t}^{\prime \, 2}}{2}+\mathcal{O}\left( \tilde{z}_{t}^{\prime \, 4}\right),$$ 
and considering a continuous time approach in eq.~(\ref{arch-def}), we might interpret $a$ as the coefficient that is
related to the magnitude of additive noise, which does not lead to ``fat tails'' in 
$p\left( z\right) $, whereas $b$ is associated with the strength of
multiplicative noise which is responsible for the emergence of tails in $%
p\left( z\right) $~\footnote{%
When the distribution for $\omega $ is non-Gaussian, $b$ answers for
the increase in the tails of $p\left( z\right) $.}~\cite{gardiner}.
% In regard of the kernel~(\ref{kernel}), index $q_{m}$ controls memory effects, 
%chiefly the influence of earlier events, more effectively than $T$.

To mathematically describe the returns probability density function we have
used the $q$-Gaussian function
\begin{equation}
p\left( z\right) =\mathcal{A}e_{q}^{-\mathcal{B}\,z^{2}},\qquad \left(
q<3\right) ,  \label{q-gaussian}
\end{equation}
with $\mathcal{B}=\left[ \bar{\sigma}_{q}^{2}\left( 3-q\right) \right] ^{-1}$%
, where, $$\bar{\sigma}_{q}^{2}\equiv \int z^{2}\left[ p\left( z\right) \right]
^{q}dz/\int \left[ p\left( z\right) \right] ^{q}dz,$$ is the $q$-generalised second order moment~\cite{3ver}, and $\mathcal{A}$ is
the normalisation constant. For $q<5/3$, $\bar{\sigma}_{q}^{2}$ relates to
the usual variance according with 
\mbox{$\bar{\sigma}_{q}^{2}\left( 3-q\right) =\bar{\sigma}^{2}\left( 5-3\,q\right) $}~\cite{ct-GM-CT}. 
%~\footnote{We have introduced a bar over 
%$\sigma $ to distinguish the local standard deviation from the standard deviation 
%obtained from the whole time series, {\i.e.} the stationary 
%standard deviation}. 
Distribution~(\ref{q-gaussian}) optimises non-additive (or Tsallis) entropy, $S_{q}$~\cite{ct}, 
and it is widely applied to describe the PDF of returns in stock market
indices and other natural and artificial processes which present the
properties that we aim to reproduce~\footnote{%
Within a financial context, distribution (\ref{q-gaussian}) is usually
referred to as $t$-Student distribution which is equivalent to the $q$%
-Gaussian distribution for $q>1$ as it can be easily checked.}. In the
characterisation of $p\left( z\right) $, all PDF adjustments have only
involved one parameter, the index $q$, since we have normalised $z$ by the standard
deviation and we have divided $p\left( z\right) $ by $p\left( 0\right) =%
\mathcal{A}$. Nevertheless, as we shall see further on, the agreement at the peak is clear-cut.

\begin{figure}[tbh]
\begin{center}  
\includegraphics[width=0.45\columnwidth,angle=0]{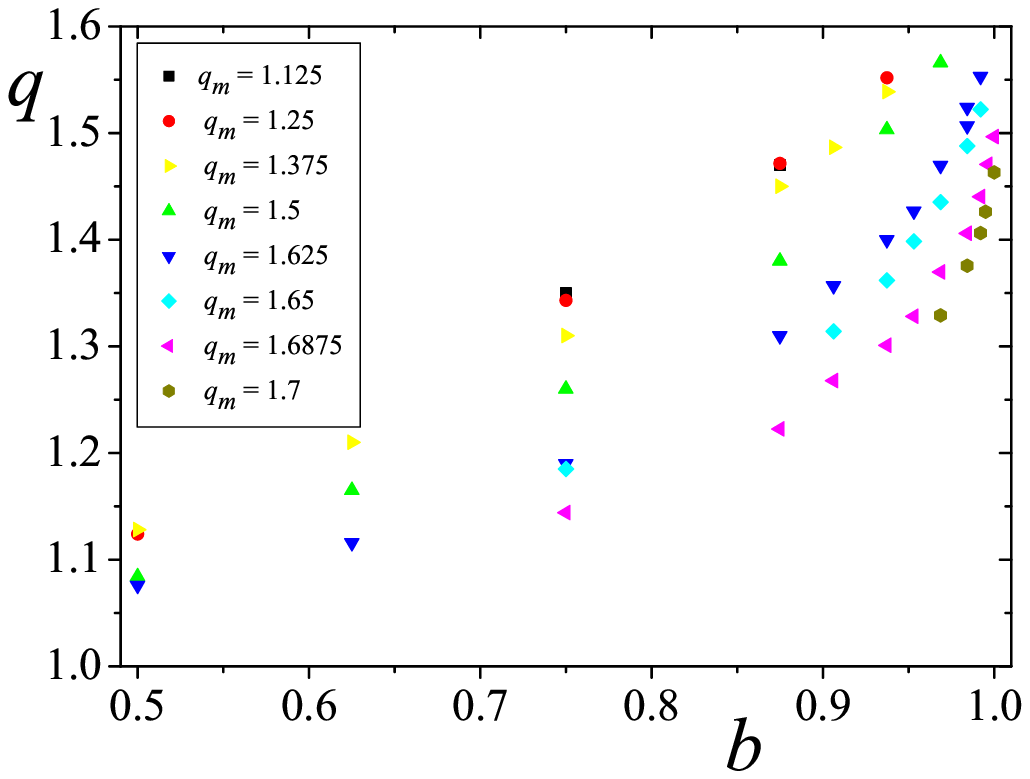} 
\includegraphics[width=0.45\columnwidth,angle=0]{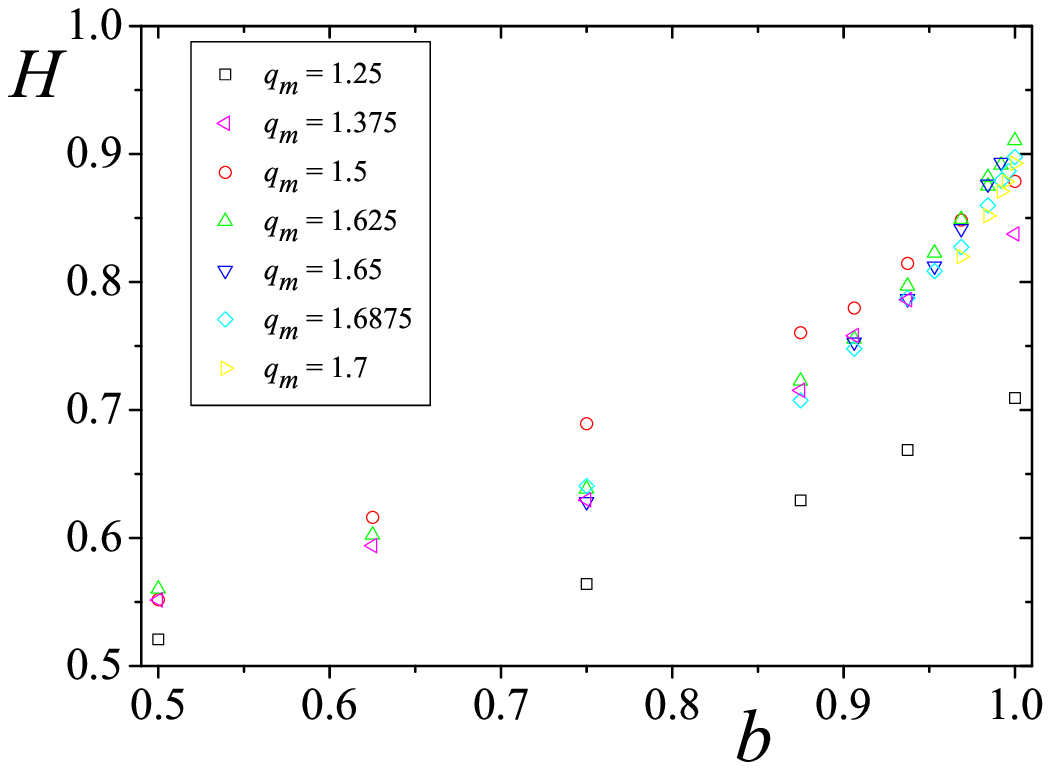}
\end{center}
\caption{
Left panel: Entropic index $q$ of eq.~(\ref{q-gaussian}) \textit{versus}
parameter $b$ for several values of memory index $q_{m}$. All the numerical
adjustments to obtain $q$ have a $\chi ^{2}$ (per degree of freedom) error
function of the order of $10^{-5}$ and squared correlation coefficient $%
R^{2}\sim 0.999$. Right panel: Hurst exponent $H$ \textit{versus} parameter $%
b$ for several values of memory index $q_{m}$ obtained by DFA method. The
numerical adjustments we have made present a correlation coefficient $R\sim
0.999$. The runs that lead to the values depicted in both panels have $%
10^{6} $ elements.
}
\label{fig-1}
\end{figure}

On account of difficulties~\footnote{%
In the evaluation of $C_{\tau }\left( x\right) $ the stationarity of the
signal is assumed, fact that does not necessarily correspond to its actual nature.
Another problem is the high sensitivity of $C_{\tau }\left( x\right) $ to the actual 
average of $x\left( t\right) $.} about evaluating truthful values of the self-correlation
function,%
\begin{equation}
C_{\tau }\left( x\right) =\frac{\left\langle x\left( t\right) \,x\left(
t+\tau \right) \right\rangle -\left\langle x\left( t\right) \,\right\rangle
^{2}}{\left\langle x\left( t\right) \,^{2}\right\rangle -\left\langle
x\left( t\right) \,\right\rangle ^{2}},  \label{correlation}
\end{equation}%
we have opted to use the integrated $\left\vert z_{t}\right\vert $ time
series Hurst exponent, $H$, obtained from the trustworthy DFA method which
describes the scaling of the root-mean square, $F\left( \tau \right) $, in
signals, $F\left( \tau \right) \,\sim \,\tau ^{H}$ ($0<H<1$)~\footnote{%
For $0<H<\frac{1}{2}$ the signal is \textit{anti-persistent} and composed by
anti-correlations, while for $\frac{1}{2}<H<1$ the time series is \textit{%
persistent} with correlations as strong as higher $H$ is. When $H=\frac{1}{2}
$ the time series is a Brownian motion (or white noise) analogue.}~\cite{dfa}. 
The results of $q$ and $H$ obtained from numerical adjustment procedures
are depicted in fig.~\ref{fig-1} as functions of $b$ and $q_{m}$.

As it is visible from fig.~\ref{fig-1} (left panel), for constant $q_{m}$, $q$
increases monotonically as $b$ also increases. For the same $b$ we observe
that larger values of $q_{m}$ lead to smaller values of $q$. In other words,
by increasing $q_{m}$, we augment memory in $\sigma _{t}^{2}$, hence
volatility tends to become less fluctuating. As a consequence, $p\left(
z\right) $ approaches $\omega $ distribution, since, as we have mentioned 
above, the time dependence of $\sigma _{t}$ is the responsible for emergence
of the tails in $p\left( z\right) $. This effect is perfectly observed when $q_{m}=\infty $, for which 
memory efects are so strong (every single element of the past influences the present with the same weight), that after some time steps volatility remains constant.

Concerning Hurst exponent figures, we would like to refer that they have a
bearing on the time interval, $\tau ^{\prime }$, before the crossover into $%
H=\frac{1}{2}$ regime. Whatever the value of $q_{m}$ we have considered, for
values of $b < 0.75$, the crossover is visible with a transition $\tau 
$, $\tau _{c}$, which increases as $b$ gets larger. Should time series have
highly persistent volatility, like price fluctuation ones, the crossover is
basically unperceptive within a temporal scale up to $\tau = 10 ^{6}$ time steps.

\medskip

\begin{figure}[tbh]
\begin{center}  
\includegraphics[width=0.45\columnwidth,angle=0]{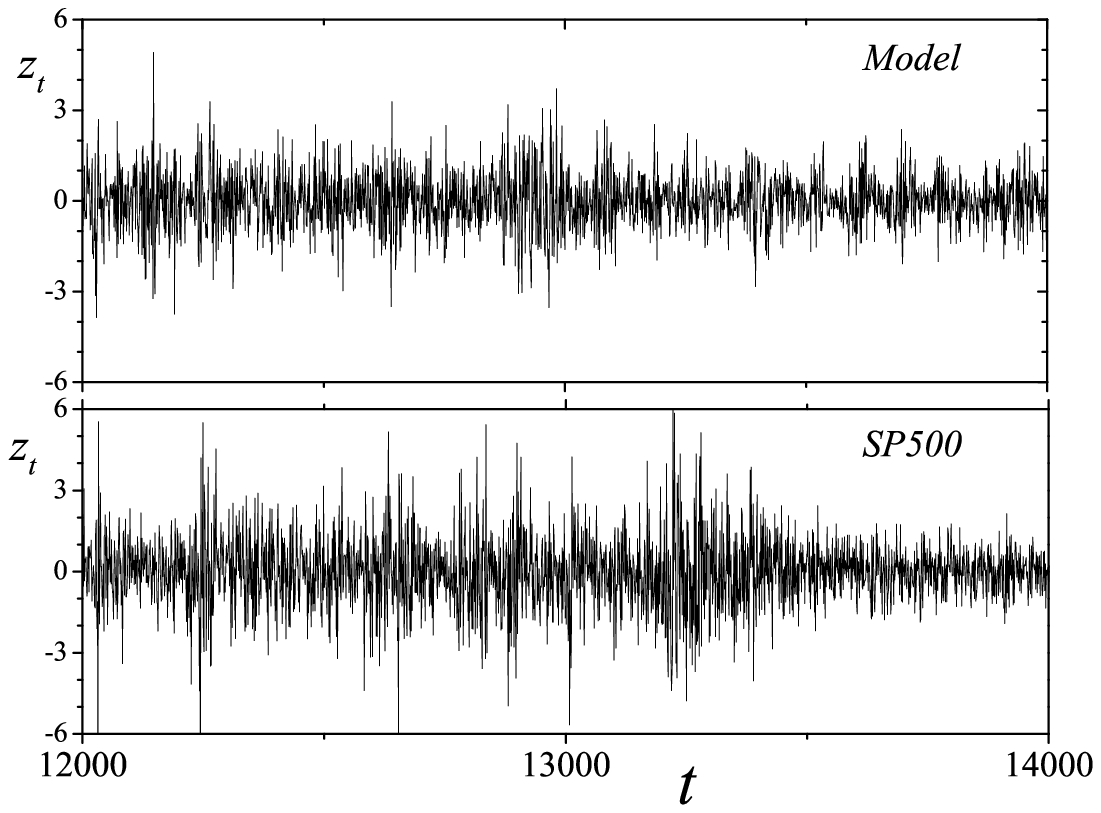} 
\includegraphics[width=0.45\columnwidth,angle=0]{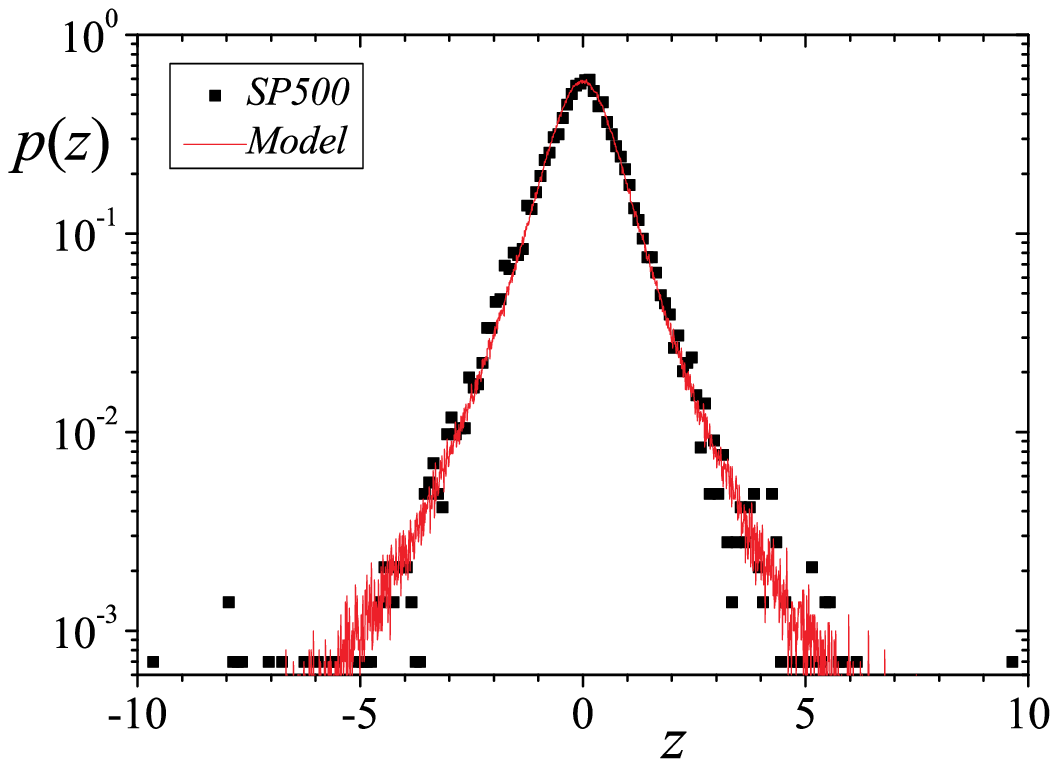}
\end{center}
\caption{
Left panels: Excerpts of $SP500$ daily normalised return times series with $2000$
elements, from the $10^{th}$ September $1997$ to the $25^{th}$ October $2005$%
, and model (with $q_{m}=1.6875$ and $b=0.99635$) for mere illustration
proposes. Right panel: Probability density function $p\left( z\right) $ 
\textit{versus} (normalised) $z$, obtained from the whole time series shown on left
panels, in $\log $-linear scale (symbols are used for $SP500$ PDF and line
for model PDF). As it can be seen the accordance is quite good. For $SP500$
ftting $q=1.47\pm 0.2$ ($\chi ^{2}=6\times 10^{-5}$ and $R^{2}=0.99$). In both cases $z$ is expressed in 
standard deviation units.
}
\label{fig-2}
\end{figure}

\begin{figure}
\includegraphics[width=0.45\columnwidth,angle=0]{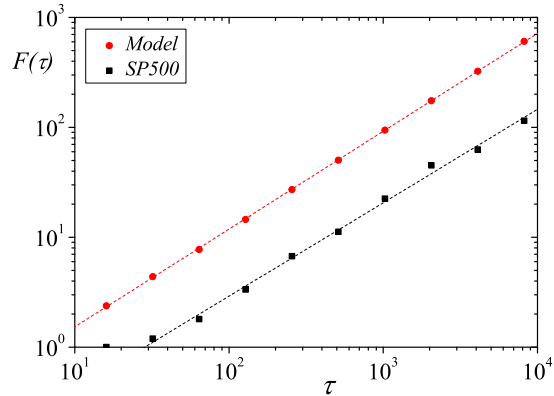}
\caption{
Root-mean square fluctuations, $F\left( \tau \right) $, \textit{%
versus} $\tau $ of the time series of fig.~\ref{fig-2}. The values obtained from
numerical fitting are $H=0.883\pm 0.005$ for $SP500$ (squares) and $%
H=0.886\pm 0.003$ for the model (circles).
}
\label{fig-3}
\end{figure}

\begin{figure}
\includegraphics[width=0.45\columnwidth,angle=0]{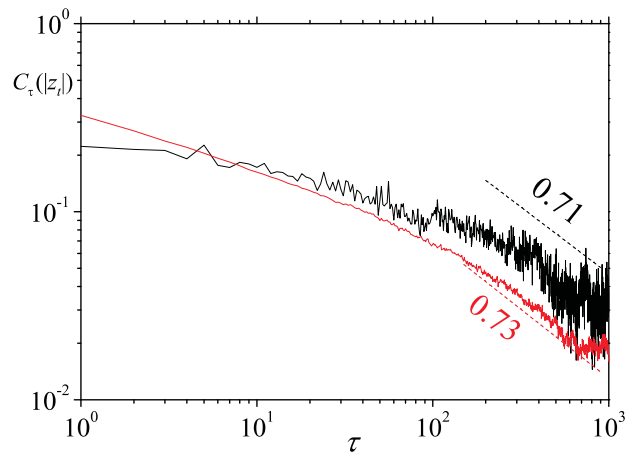}
\caption{
Absolute returns self-correlation function $C_{\tau }\left( \left\vert z_{t}\right\vert
\right) $ \textit{versus} $\tau $ in $\log $-$\log $ scale. The similarity
on the qualitative behaviour of the two curves is evident. Although the fluctuation, evaluating the
decay exponents as a result of a fitting procedure for large $\tau $ we have obtained $0.73\pm 0.01$ for the
model and $0.71\pm 0.03$ for $SP500$.
}
\label{fig-4}
\end{figure}

From the set of numerical results we have estimated the best values of $q_{m}
$ and $b$ which can reproduce statistical features of a \textit{paragon} of
the type of time series which we have been referring to --- the daily fluctuations of $SP500
$ financial index~\cite{econofisica}. Our $SP500$ time series runs from the $3^{rd}$ January $%
1950$ up to the $28^{th}$ February $2007$ in a total of $14380$ business days. The
daily return $z_{t}$ is computed as $$z_{t} \equiv \ln \,S_{t}-\ln \,S_{t-1},$$ where 
$S_{t}$ represents the $SP500$ value at time $t$. As it is usual we have didived $z$ by its standard deviation. Gathering together the values
of $q$ and $H$ for $SP500$, respectively $1.47\pm 0.02$ and $0.88$, we have
verified that $q_{m}=1.6875$ and $b=0.99635$ are able to reproduce, with a
remarkable agreement, both the return probability density function and the
Hurst exponent as it is exhibited on fig.~\ref{fig-2} and fig.~\ref{fig-3}. Further, when we have
compared, \emph{a posteriori}, the self-correlation functions of $\left\vert
z_{t}\right\vert $, eq.~(\ref{correlation}), we have verified the same
qualitative behaviour. In fact, despite both of the short range available for fitting and the fluctuations, 
a quite similar power-law decay with an
exponent of $0.73\pm 0.01$ for our model and $0.71\pm 0.02$ for $SP500$ as
shown on fig.~\ref{fig-4}. Specifically, and according to fig.~\ref{fig-4}, the two curves
stand basically side by side for $\tau > 20$ in a $\log -\log $ scale.

\medskip

To summarise, in this manuscript we have introduced a generalisation of 
\textsc{Engle}'s proposal for generating instantaneous volatility in
heteroskedastic processes. This modification refers to the introduction of a
memory kernel which has an asymptotic power-law dependence defined by a 
parameter $q_{m}$. Apart from the fact that our alteration has been able to reobtain the
non-Gaussian PDF for the random variable, $z_{t}$, it has also been successful
about reproducing the long-lasting (asymptotic power-law decaying)
self-correlation of the magnitude of $z_{t}$ exhibited by a large number of
phenomena. The improvement in the reproduction of statistical features of
such a kind of time series has been achieved by considering just one
additional parameter, $q_{m}$, which represents a clear simplification
against $ARCH\left( s\right) $ (with $s\gg 1$), that only manage to exhibit a exponential
volatility self-correlation function with large characteristic time, or other heteroskedastic processes~\cite{arch-rev}. By
exhaustive numerical analysis of our model we have found a pair of values, $%
q_{m}$ and $b$, with which we have mimicked daily fluctuations of $SP500$.
The resemblance between $SP500$ time series and the signal obtained by
numerical application of our suggestion is remarkably good for the
probability density function and the Hurst exponent. In a qualitative sense,
the correlation function has also been quite well described. The
quantitative discrepancies verified in $C_{\tau }\left( \left\vert z_{t}\right\vert
\right) $ and $F\left( \tau \right) $ might be solved if we modify kernel~(\ref{kernel}) 
by introducing a sort of ``characteristic time'', $T$,{\it i.e.}, in eq.~(\ref{kernel}) $t^{\prime} \rightarrow t^{\prime}/T$, as another parameter.

It is well known that there are an infinity of dynamics whose outcome is the same probability density function. However, 
as far as we are able to obtain an appropriate reproduction of further statistical properties, as it is the case we have just presented, 
we will be approaching our models towards the nature of the system upon study. This is certainly important when the models are applied, {\it e.g.}, 
on forecasting purposes. It is on this basis we support the relevance of our propose.

In respect of financial markets, and considering a macroscopic
approach, our model permit us to say that price fluctuations are actually
dependent on their history, but on a asymptotically scale-free way~\cite{tsallis-ca}, 
as it is exhibited by the majority of the so-called {\it complex systems}. Such a dependence is in contrast with the usual, 
and analitycally simpler, exponential treatment. 
On a practical way, this also means that past events take long time to loose their importance.

Last of all, owing to $C_{\tau }\left( \left\vert z_{t}\right\vert \right) $
asymptotic power-law decay, as it is visible from eq.~(\ref{q-exp}), we
could make a correspondence between the decay exponents and a correlation index, $q_{c}$. By this
we get $q_{c}=2.37\pm 0.03$ for our model and $q_{c}=2.41\pm 0.06$ for $SP500$. 
Such an association introduces an alternative triplet of entropic
indices~\cite{tsallis-villa}, namely $\left\{ q_{m},q,q_{c}\right\} $,
related to non-extensive statistical mechanics formalism that could
characterise this type of systems.

\bigskip

SMDQ acknowledges \textsc{C. Tsallis} for his continuous encouragement and discussions as well as \textsc{E.~M.~F. Curado} for 
helpful and stimulating conversations at early and final stages of the work. \textsc{E.~P. Borges} and \textsc{F.~D. Nobre} are thanked for comments made on 
previous versions of this manuscript. This work has benefited from infrastructural
support from PRONEX/MCT (Brazilian agency) and financial support from FCT/MCES (Portuguese agency).

\end{document}